\documentclass[english,reprint,aps,superscriptaddress]{revtex4-1}
\usepackage[T1]{fontenc}
\usepackage[latin9]{inputenc}
\usepackage{geometry}
\usepackage{babel}
\usepackage{amsmath}
\usepackage{amssymb}
\usepackage{graphicx}
\usepackage[unicode=true]
 {hyperref}



\begin{document}
\title{Signature of Parity Anomaly: Crossover from One Half to Integer Quantized
Hall Conductance in a Finite Magnetic Field}
\author{Huan-Wen Wang}
\affiliation{School of Physics, University of Electronic Science and Technology
of China, Chengdu 611731, China}
\author{Bo Fu}
\affiliation{School of Sciences, Great Bay University, Dongguan 523000, China}
\author{Shun-Qing Shen}
\email{sshen@hku.hk}

\affiliation{Department of Physics, The University of Hong Kong, Pokfulam Road,
Hong Kong, China}
\begin{abstract}
The pursuit of understanding parity anomaly in condensed matter systems
has led to significant advancements in both theoretical and experimental
research in recent years. In this study, we explore the parity anomaly
of massless Dirac fermions in a semimagnetic topological insulator
(TI) thin film subjected to a finite magnetic field. Our findings
reveal an anomalous half-quantized Hall conductance arising from the
occupied electronic states far below the Fermi level, which is directly
associated with the parity anomaly. This observation demonstrates
a crossover from one-half quantized Hall conductance in a metallic
phase at zero field to one or zero quantized Hall conductance in the
insulating phase at a strong field in the presence of disorders, serving
as a key indicator for confirming parity anomaly. Our work provides
valuable insights into the intricate relationship between band topology
in condensed matter systems and quantum anomaly in quantum field theory.
\end{abstract}
\maketitle

\paragraph*{Introduction}

Parity anomaly is a conflict between the parity symmetry and U(1)
gauge invariance of massless Dirac fermions in quantum field theory
\citep{Niemi1983axial,SemenoffPRL1984,Redlich1984gauge,Matthew-19prb,FradkinPRL1986},
and exhibits a half-quantized Hall conductance in transport measurements
\citep{Jackiw1984Fractiona,Boyanovsky1986physical,Schakel1991relativistic,Haldane1988Model,Mogi-21np,FuB-22xxx,Hu-2022prb,Zou-22prb,zou-23prb}.
Up to now, several condensed matter systems have been proposed to
realize parity anomaly \citep{FradkinPRL1986,Haldane1988Model,SemenoffPRL1984}.
In his seminal paper \citep{Haldane1988Model}, Haldane pointed out
that massless Dirac fermions occur along the critical line between
Chern insulators and trivial insulators. Besides, the massless Dirac
fermions are also found on the boundary of three-dimensional TIs \citep{Fu2007topological,hasan2010colloquium,Qi2008topological,shen2012topological,XuY-14np,Chu-11prb,Koenig2014half,Zhang2017anomalous,chang23rmp}.

\begin{figure}
\centering{}\includegraphics[width=7.5cm]{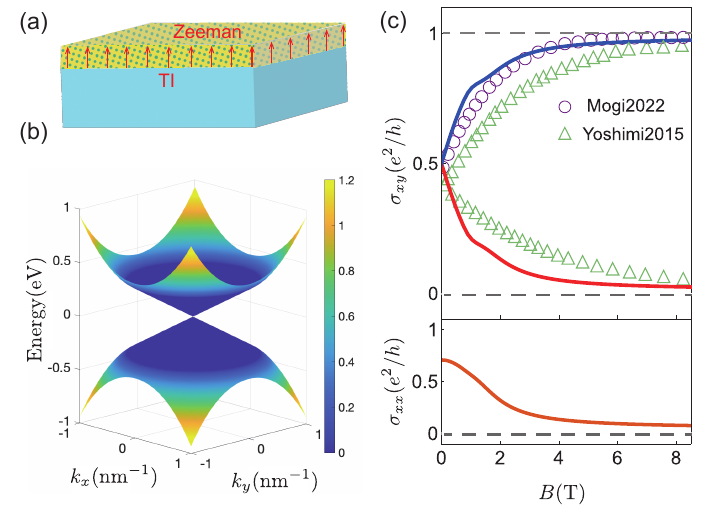}\caption{\label{fig:Schematic}(a) Schematic of a semi-magnetic TI film. (b) The massless Dirac cone in a two-dimensional Brillouin zone, the color
represents the relative Dirac mass $m(k^{2})/v\hbar k$, in the parity
symmetry invariant regime, $m(k^{2})/v\hbar k=0$. (c) The upper panel, crossover from half- to integer-quantized Hall conductance in a finite magnetic field. The open markers are the experimental data extracted from Refs. \citep{Mogi-21np,Yoshim-15nc}. The lower panel, longitudinal conductance in a finite magnetic field, which is finite at $B=0$, and tends to zero in the strong magnetic field. Here the calculation parameters are chosen as $v\hbar=410\,\mathrm{meV\cdot nm}$, $b\hbar^{2}=-566\,\mathrm{meV}\cdot\mathrm{nm}^{2}$, $k_{c}=0.7\,\mathrm{nm}^{-1}$, $\kappa=0.001\,\mathrm{nm}^{-2}$.}
\end{figure}

Jackiw \citep{Jackiw1984Fractiona} firstly predicted the one-half
quantized Hall conductance of massless Dirac fermions in a continuous
model at a finite magnetic field, in which the sign of the Hall conductance
depends on whether the zero mode of the Landau levels is filled or
not. This prediction was extensively accepted after the observation
of quantum Hall effect in graphene, in which the Hall conductance
was observed to be $\sigma_{xy}=2(2\nu+1)\frac{e^{2}}{h}$ ($\nu$
is an integer) \citep{Novoselov-05nature,Zhang-05nature,Neto-09-rmp}.
Considering the double spin and valley degeneracy, it was reasonably
assumed that each Dirac cone contributes $(\nu+\frac{1}{2})\frac{e^{2}}{h}$
to the Hall conductance \citep{Gusynin-05prl}. Besides, a strong
TI thin film also hosts a pair of the massless Dirac cones of surface
fermions \citep{Fu2007topological,hasan2010colloquium,Qi2008topological,shen2012topological,XuY-14np}.
For the non-doped case, each Dirac cone is expected to contribute
$(\nu+\frac{1}{2})\frac{e^{2}}{h}$ to the Hall conductance at a finite
magnetic field \citep{XuY-14np,Yoshimi15nc,Li17prb,chong19prl}, which
is similar to graphene. By magnetically doping on the one side of
a film, or in proximity to a magnetic insulating layer as illustrated
in Fig. \ref{fig:Schematic}(a), one surface state gap out \citep{liu09prl,Chen-10science,Chang-13science,Tokura2019MTI},
allowing the formation of a single flavor of massless Dirac fermions
on the TI film as schematically shown in Fig. \ref{fig:Schematic}(b)
\citep{Yoshim-15nc,Mogi-21np,lu21prx}. Recently, the measurement
of the half-quantized Hall conductance at zero magnetic fields was
reported in such a semi-magnetic TI \citep{Mogi-21np}, and several
mechanisms have been proposed \citep{Zou-22prb,Zhou2022prl,zou-23prb,Gong2023nsr}.
Meanwhile, the conductance flow with decreasing temperatures was shown
to converge to the point of $\sigma_{xx}=0$ or $\sigma_{xy}=\frac{e^{2}}{h}$
at a magnetic field of $9\mathrm{T}$, indicating that an integer
quantized Hall conductance occurs for a single flavor of massless
Dirac fermions at a finite field as shown in Fig. \ref{fig:Schematic}(c).
This is obviously opposite to Jackiw's prediction, and inconsistent
with experimental observation in graphene and non-magnetic TI films.
The flow diagram also shows a qualitatively different pattern to the
conventional Hall effect, which causes some confusion in the community
\citep{Beennakker2022}.

In the present work, we investigate the Hall conductance for a single Dirac cone of the surface fermions in a magnetic field at a semimagnetic TI film by including a time-reversal symmetry breaking term at the
higher energy part, which is necessary for the realization of a single flavor of massless Dirac fermions on a lattice. We also demonstrate the evolution of the Hall conductance from one-half at zero fields to zero
or one at strong fields in the presence of disorder as indicated by the solid lines in the upper panel of Fig. \ref{fig:Schematic}(c). The evolution illustrates that the parity anomaly-related physics occurs for a system with a single flavor of massless Dirac fermions, which is absent in the system with a pair of massless Dirac fermions such as in graphene and TI films. The experimental data {[}open markers in Fig. \ref{fig:Schematic}(c){]} extracted from two different and independent experiments on semi-magnetic TIs \citep{Mogi-21np,Yoshim-15nc} have already revealed some signatures of parity anomaly in a finite magnetic field. This work clarifies the topological difference between the systems with a single flavor and two distinct flavors of massless Dirac fermions.

\paragraph*{Model Hamiltonian}

The gapless surface electrons in a three-dimensional TI are usually
modeled as ideal Dirac fermions with a linear dispersion \citep{Shan2010njp}.
However, when dealing with a single flavor of massless Dirac fermions
on a lattice, the inclusion of a time-reversal symmetry broken term
in the dispersion becomes inevitable in order to avoid the violation
of the fermion doubling theorem \citep{Wilson-75,Rothe,Nielsen1981PLB}.
Here we start with an effective model for a single massless Dirac
cone of the surface states subjected to a perpendicular magnetic field
\citep{zou-23prb}
\begin{equation}
H=\begin{pmatrix}M(\Pi^{2}) & v(\Pi_{x}-i\Pi_{y})\\
v(\Pi_{x}+i\Pi_{y}) & -M(\Pi^{2})
\end{pmatrix}\label{eq:hamiltonian}
\end{equation}
where the kinematic momentum operator $\Pi_{\alpha}=-i\hbar\partial_{\alpha}-eA_{\alpha}$.
The momentum-dependent mass term $M(\Pi^{2})=f\left(k_{c}^{2}-\hbar^{-2}\Pi^{2}\right)m\left(\Pi^{2}\right)$,
$m(\Pi^{2})=-b\left(\Pi^{2}-\hbar^{2}k_{c}^{2}\right)$ and $f(x)=[1+\exp(x/\kappa)]$
is the Fermi-Dirac-distribution-like factor with a tiny constant $\kappa$.
$\hbar k_{c}$ is a material-specific parameter around which the surface
states evolve into the bulk state at a higher momentum part. The model contains the massless linear dispersion at low energy, and also a term that breaks the time-reversal symmetry at higher energy, which is necessary for the existence of a single Dirac cone on a lattice. The linear term alone does not reflect the anomaly-related physics of the single flavor of massless Dirac fermions\citep{zou-23prb}. The model can be derived analytically for the surface electrons from a three-dimensional model for TI.

\begin{figure}
\centering{}\includegraphics[width=7.5cm]{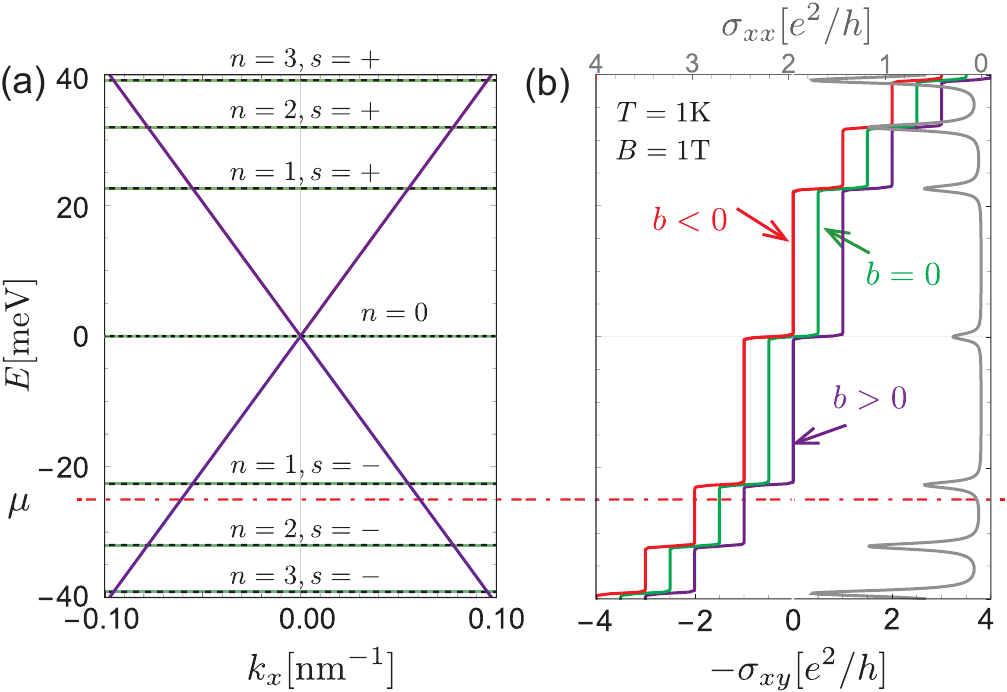}\caption{\label{fig:Energy-dependence-of}(a) The energy spectrum in the parity
symmetry invariant regime. The flat bands are the Landau levels at
$B=1\,\mathrm{T}$, and the purple line is the energy dispersion in
the zero magnetic field. (b)Energy dependence of Hall conductances in
Eq. (\ref{eq:Hall-clean}) and longitudinal conductance (the gray
line) for $b\hbar^{2}=-566\,\mathrm{meV}\cdot\mathrm{nm}^{2}$, $0$
and $566\,\mathrm{meV}\cdot\mathrm{nm}^{2}$ at $B=1\,\mathrm{T}$
and $T=1\,\mathrm{K}$. The other calculation parameters are chosen
as $v\hbar=410\,\mathrm{meV\cdot nm}$, $k_{c}=0.7\,\mathrm{nm}^{-1}$,
$\kappa=0.001\,\mathrm{nm}^{-2}$. For the longitudinal conductance
$\delta=0.5\,\mathrm{meV}$.}
\end{figure}

For a uniform magnetic field $B$ along the $z$-direction, we take the gauge field as $\mathbf{A}=(-By,0,0)$. In this case $-i\partial_{x}$ commutes with the Hamiltonian and can be replaced by its eigenvalues $k_{x}$. The ladder operators are introduced as $a=((y_{0}-y)/\ell_{B}-\ell_{B}\partial_{y})/\sqrt{2}$ and $a^{\dagger}=[(y_{0}-y)/\ell_{B}+\ell_{B}\partial_{y}]/\sqrt{2}$ \citep{shen-05prb} where the magnetic length $\ell_{B}=\sqrt{eB/\hbar}$ and the guiding center $y_{0}=\hbar k_{x}/eB$. Then the Hamiltonian is reduced to
\begin{equation}
H=\begin{pmatrix}M[\frac{\hbar^{2}}{\ell_{B}^{2}}(2a^{\dagger}a+1)] & \eta a\\
\eta a^{\dagger} & -M[\frac{\hbar^{2}}{\ell_{B}^{2}}(2a^{\dagger}a+1)]
\end{pmatrix}\label{eq:a-form}
\end{equation}
with the cyclotron energy $\eta=\sqrt{2}v\hbar/\ell_{B}$. Take the ansatz for the energy eigenstate $|\psi\rangle=(a_{n}\left|n-1\right\rangle ,b_{n}\left|n\right\rangle )^{T}$ ($T$ represents transpose) and $\left|n\right\rangle $ is the eigenket of $a^{\dagger}a$ with an integer eigenvalue $n$. The energy eigenvalues are
\begin{align}
\varepsilon_{ns} & =\frac{\omega_{n}}{2}+s\sqrt{m_{n}^{2}+n\eta^{2}},\label{eq:energy_spectrum-1}
\end{align}
where $\omega_{n}=M[\frac{\hbar^{2}}{\ell_{B}^{2}}(2n-1)]-M[\frac{\hbar^{2}}{\ell_{B}^{2}}(2n+1)]$
and $m_{n}=\frac{1}{2}\left\{ M[\frac{\hbar^{2}}{\ell_{B}^{2}}(2n-1)]+M[\frac{\hbar^{2}}{\ell_{B}^{2}}(2n+1)]\right\} $
with $s=\pm$ for $n\ge1$ and $s=-1$ for $n=0$. For $n>n_{c}=b\hbar^{2}k_{c}^{2}/\omega,$
$\varepsilon_{ns}=\frac{\omega}{2}+s\sqrt{(n-n_{c})^{2}\omega^{2}+n\eta^{2}}$
with $\omega=2b\hbar^{2}/\ell_{B}^{2}$. For $n<n_{c}$, $\varepsilon_{ns}=s\sqrt{n}\eta$
which are the Landau spectra of massless Dirac fermions \citep{Novoselov-05nature,shen2012topological}.
The Landau degeneracy per unit area for each Landau level is $n_{L}=eB/h$.
We have plotted the Landau spectra in the low energy in Fig. \ref{fig:Energy-dependence-of}(a).

\paragraph*{Kubo-Streda formula}

In general, the total Hall conductivity for a two-dimensional electron system can be evaluated by means of the Kubo-Streda formula \citep{streda-1982,wang-19prb}
\begin{align}
\sigma_{xy}= & \mathrm{Im}\frac{e^{2}\hbar}{\pi\Omega}\sum_{k}\int_{-\infty}^{+\infty}n_{F}(\epsilon-\mu)d\epsilon\nonumber \\
 & \times\text{Tr}\left[\hat{v}^{x}\frac{dG^{R}}{d\epsilon}\hat{v}^{y}\mathrm{Im}G^{R}-\hat{v}^{x}\mathrm{Im}G^{R}\hat{v}^{y}\frac{dG^{A}}{d\epsilon}\right],\label{eq:kubo-streda}
\end{align}
where $\Omega$ is the area of the two dimensional system, $G^{R/A}=[\epsilon-H\pm i\delta]^{-1}$
is the retarded or advanced Green's function, $\delta$ is the disorder-induced band broadening, $\hat{v}^{x}=i\hbar^{-1}[H,x]$ and $\hat{v}^{y}=i\hbar^{-1}[H,y]$ are the velocity operators along the $x-$ and $y-$direction, respectively. $n_{F}(\epsilon-\mu)=\left[1+\exp\left(\frac{\epsilon-\mu}{k_{B}T}\right)\right]{}^{-1}$ is the Fermi-Dirac distribution function with $\mu$ the chemical potential and $k_{B}T$ the product of Boltzmann constant and temperature. In this work, we will take advantage of Eq. (\ref{eq:kubo-streda})
to study the anomaly-related Hall conductance for both disordered and clean systems.

In the zero magnetic field and clean limit, Eq. (\ref{eq:kubo-streda}) leads to the anomalous Hall conductance for massless Dirac fermions as \citep{note-SI}
\[
\sigma_{xy}^{A}=-\frac{e^{2}}{2h}\left[\mathrm{sgn}(b)+\frac{M(\hbar^{2}k_{F}^{2})}{|\mu|}\right],
\]
where $k_{F}$ is the Fermi wave vector and can be found by solving $\mu^{2}=v^{2}\hbar^{2}k_{F}^{2}+M^{2}(\hbar^{2}k_{F}^{2})$. In the parity-symmetry invariant regime, $k_{F}<k_{c}$ and $M(\hbar^{2}k_{F}^{2})=0$, the Hall conductance is half-quantized as $\sigma_{xy}^{A}=-\frac{e^{2}}{2h}\mathrm{sgn}(b)$, which is only determined by the sign of the quadratic term $m(\Pi^{2})$
at higher energy \citep{zou-23prb}. It can be regarded as a transport signature of parity anomaly in the absence of a magnetic field. For $k_{F}>k_{c}$, $M(\hbar^{2}k_{F}^{2})\ne0$, an additional correction
makes the Hall conductance deviating from the half-quantized value.

\paragraph*{Integer-quantization and parity anomaly}

To reveal the consequence of the parity anomaly of massless Dirac fermions at a finite magnetic field, we begin with an ideal case of $\delta=0$. By utilizing the eigenstates and eigenvalues of Eq. (\ref{eq:a-form}),
we can calculate the retarded and advanced Green's function explicitly. Substituting the calculated Green's functions into the Kubo-Streda formula, one can obtain the Hall conductance at a finite temperature $T$ and chemical potential $\mu$ as \citep{note-SI}
\begin{equation}
\sigma_{xy}=-\frac{e^{2}}{2h}\mathrm{sgn}(b)+\sigma_{xy}^{0}\label{eq:Hall-clean}
\end{equation}
with 
\[
\sigma_{xy}^{0}=\frac{e^{2}}{2h}\left[1-2\sum_{ns}sn_{F}(s\varepsilon_{ns}-s\mu)\right]
\]
where $s=\pm1$ for $n>0$ and $s=1$ for $n=0$. $\sigma_{xy}^{0}$ has the identical form to the Hall conductance of ideal massless Dirac fermions \citep{Gusynin-05prl} and is half-quantized as indicated
by the green line in Fig. \ref{fig:Energy-dependence-of}. The term $\frac{e^{2}}{2h}$ in $\sigma_{xy}^{0}$ is caused by the fact that the lowest Landau level of Dirac fermion has twice smaller degeneracy
than the levels with $n>0$. $-\frac{e^{2}}{2h}\mathrm{sgn}(b)$ in Eq. \ref{eq:Hall-clean} is an additional contribution from the high energy part as $\lim_{n\to+\infty}\frac{m_{n}}{E_{n}}=-\mathrm{sgn}(b)$
and is a manifestation of parity anomaly. As indicated by the red ($b<0$) and purple ($b>0$) solid lines in Fig. \ref{fig:Energy-dependence-of}, the existence of parity anomaly ($-\frac{e^{2}}{2h}\mathrm{sgn}(b)$)
restores the integer-quantization of Hall conductance once the chemical potential is inside the gap of two different Landau levels. From the bulk-edge correspondence, the integer quantum Hall conductance corresponds to the number of edge states. Moreover, as shown by the gray solid
line in Fig. \ref{fig:Energy-dependence-of}(c), the longitudinal conductance $\sigma_{xx}$ is independent of the $b$ term in the low energy window, and it becomes zero when the Hall conductance is
quantized, which makes the measurement more accessible in experiments.

The appearance of $-\frac{e^{2}}{2h}\mathrm{sgn}(b)$ can also be understood from the spectral asymmetry or Atiyah-Patodi-Singer eta invariant $\eta_{H}=\sum_{n,s,k_{x}}\mathrm{sgn}(\varepsilon_{ns})$ \citep{APS-index,niemi-eta}. In general, $\eta_{H}$ is not convergent
and required to be regulated. Here we apply a heat-kernel regularization, i.e., $\eta_{H}=\lim_{\kappa\to0^{+}}\text{\ensuremath{\sum}}_{k_{x},ns}\mathrm{sgn}(\varepsilon_{ns})e^{-\kappa|\varepsilon_{ns}|}$ \citep{niemi-eta,B=0000F6ttcher2019survival,B=0000F6ttcher2020fate}.
As the contribution from $\varepsilon_{n+}$ and $\varepsilon_{n-}$ is always canceled by each other for a finite $n$ except for $n=0$, we only need to consider the contribution from a large $n$. In the
case, $\omega_{n}=\omega$. We can expand the energy spectrum as $\varepsilon_{ns} \approx\frac{\omega}{2}+sn|\omega|+s\mathrm{sgn}(b)\left(\frac{\eta^{2}}{2\omega}-b\hbar^{2}k_{c}^{2}\right)$. Plugging this expression into $\eta_{H}$ and making summation over $n$ and $s$, we have \citep{note-SI}
\begin{align}
\eta_{H}= & \Omega\frac{eB}{2\pi\hbar}\left[\mathrm{sgn}(\varepsilon_{0})-\mathrm{sgn}(b)\right].\label{eq:spectral asymmetry}
\end{align}
The first term $\mathrm{sgn}(\varepsilon_{0})$ is attributed by the zeroth Landau level $(n=0)$ or low energy, and another term $\mathrm{sgn}(b)$ comes from the part of high energy. For $n=0$, the energy of the lowest Landau level is either positive or negative depending on the magnetic field and the model parameters, and it naturally contributes to the spectral asymmetry as $\frac{\Omega eB}{2\pi\hbar}\mathrm{sgn}(\varepsilon_{0})$. However, for $n>0$, since there are always two energy levels $\varepsilon_{n+}$ and $\varepsilon_{n-}$, one expects that there is no contribution to the spectral symmetry $\eta_{H}$. Hence, the appearance of $\mathrm{sgn}(b)$ in $\eta_{H}$ is abnormal, it is a consequence of the parity-symmetry breaking from the electrons of high energy. Furthermore, according to the Streda formula $\sigma_{xy}=-e\frac{\partial\varrho}{\partial B}\mid_{\mu}$ with $\varrho=-\frac{\eta_{H}}{2\Omega}+\varrho_{0}$ the number density of fermions and $\varrho_{0}$ the density of charge carriers for empty or filled states relative to the charge neutrality, the spectral asymmetry from the part of high energy corresponds to a half-quantized Hall conductance $-\frac{e^{2}}{2h}\mathrm{sgn}(b)$. Therefore, the parity-symmetry breaking in the region of high energy indeed leads to a half-quantized Hall conductance.

\paragraph*{Crossover from half- to integer-quantization}

\begin{figure}
\centering{}\includegraphics[width=7.5cm]{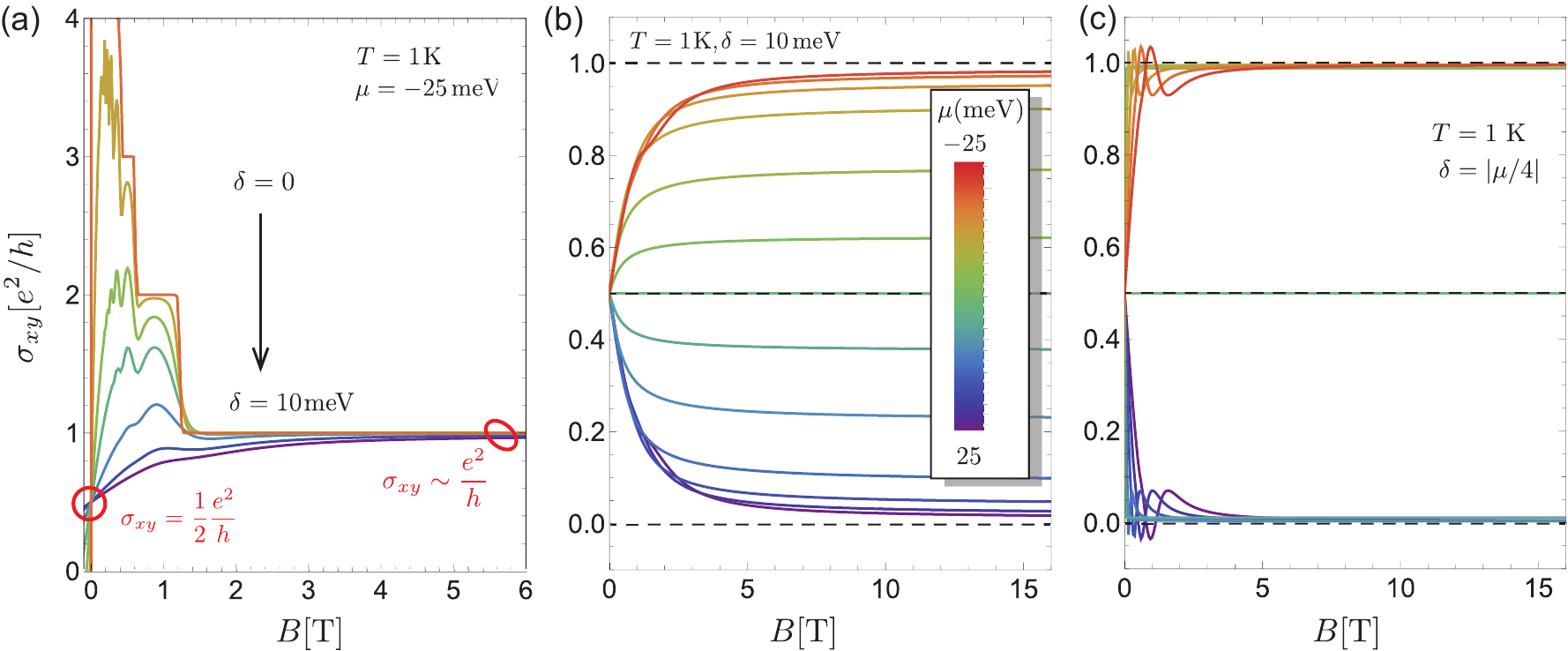}\caption{\label{fig:Magnetic-field-dependence}(a) Magnetic field dependence of Hall conductance for massless Dirac fermion with $b\hbar^2=-566\,\mathrm{meV\cdot nm}^2$ for several different band broadenings $\delta=0,1,2,3,5,8,10\,\mathrm{meV}$ in the parity symmetry invariant regime at $T=1.0\,\mathrm{K}$. Here we put the chemical potential at $\mu=-25\,\mathrm{meV}$ {[}as indicated by the red dashed line in Fig. (\ref{fig:Energy-dependence-of}){]}. (b-c) Crossover from one half to integer quantized Hall conductance for
massless Dirac fermion in the parity symmetry invariant regime. Here we choose the chemical potentials as $\mu=0,\pm2,\pm5,\pm10,\pm15,\pm20,\pm25\,\mathrm{meV}$, temperature as $T=1.0\,\mathrm{K}$, band broadening as $\delta=10\,\mathrm{meV}$ in (b) and $\delta=|\mu/4|$ in (c).}
\end{figure}

To show the crossover from the half-quantized Hall conductance at
a zero field to an integer-quantized Hall conductance at a strong field
in Ref. \citep{Mogi-21np}, we calculate the Hall conductance in the
presence of disorders according to Eq. (\ref{eq:kubo-streda}). As
displayed by the red line in Fig. \ref{fig:Magnetic-field-dependence}(a),
the Hall conductance is integer-quantized in the clean limit ($\delta\to0$),
which is consistent with Eq. (\ref{eq:Hall-clean}). When $\delta$
increases, the peak height of $\sigma_{xy}$ is suppressed as shown
in Fig. \ref{fig:Magnetic-field-dependence}(a). When $\delta$ is
strong enough, $\sigma_{xy}$ exhibits a crossover from one half-
to integer-quantization with increasing the magnetic field {[}purple
solid line in Fig. \ref{fig:Magnetic-field-dependence}(a){]}. It
should be emphasized that the Hall conductance is always half-quantized
in the absence of a magnetic field once the parity symmetry is preserved
near the Fermi surface although it is broken as a whole \citep{zou-23prb}.
Hence we have $\sigma_{xy}(0)=-\frac{e^{2}}{2h}\mathrm{sgn}(b)$ at
a finite Fermi energy in Fig. \ref{fig:Magnetic-field-dependence}(a).
In Fig. \ref{fig:Magnetic-field-dependence}, we put the chemical
potential in the valence band ($\mu=-25\,\mathrm{meV}$), $\sigma_{xy}$
changes from $0.5$ to $1$. If we put the chemical potential in the
conduction band $(\mu=25\,\mathrm{meV})$, as depicted in Fig. \ref{fig:Magnetic-field-dependence}(b),
the crossover will take place between $0.5$ and $0$, which is in
a good agreement with the experiment data by tuning the gate voltage
(see Fig. 2b in Ref. \citep{Yoshim-15nc}). If the chemical potential
is comparably even smaller than the band broadening $\delta$, the
Hall conductance will deviate from its quantized value. As shown in
Fig. \ref{fig:Magnetic-field-dependence}(b), we tune the chemical
potential from $-25\,\mathrm{meV}$ to $25\,\mathrm{meV}$ and fix
the band broadening as $\delta=10\,\mathrm{meV}$, the Hall conductance
at a strong field changes from $1$ to $0$, and is half-quantized
for $\mu=0$. In experiments, a high-quality sample ($\mu>\delta$)
is required to observe the integer-quantized Hall conductance in a
finite magnetic field. For instance, as shown in Fig. \ref{fig:Magnetic-field-dependence}(c),
if we set $\delta=|\frac{\mu}{4}|$, the Hall conductance approaches
to $1$ or $0$ for all the chosen chemical potentials except $\mu=0$.
At $\mu=0$, the Hall conductance is half-quantized as $\sigma_{xy}=-\frac{e^{2}}{2h}\mathrm{sgn}(b)$,
which is robust to the magnetic field, disorder and temperature. Meanwhile,
the longitudinal conductivity is also finite at zero temperature,
$\sigma_{xx}(\mu=0)=\frac{1}{\pi}\frac{e^{2}}{h}$ \citep{note-SI}.
It is a typical behavior of parity anomalous semimetal in the magnetic
field \citep{FuB-22xxx}.

\paragraph*{Impact of particle-hole symmetry breaking}

In the previous discussion, the Dirac fermions in Eq. (\ref{eq:hamiltonian}) preserve the particle-hole symmetry. However, the surface electrons are generally particle-hole asymmetric in TIs. If we further include additional term $-d\Pi^{2}$ in Eq. (\ref{eq:hamiltonian}), the anomalous Hall conductance becomes $\sigma_{xy}^{A}=-\frac{e^{2}}{2h}[\mathrm{sgn}(b)+\frac{M(\hbar^{2}k_{F}^{2})}{|\mu+d\hbar^{2}k_{F}^{2}|}]$ at the zero magnetic field. As the additional term is parity symmetry invariant, $\sigma_{xy}^{A}$ is still half-quantized for $k^{2}<k_{c}^{2}$. In a finite magnetic field, the crossover from half- to integer quantization is preserved. In the clean limit, the Hall conductance has an identical form as Eq. (\ref{eq:Hall-clean}) by replacing the energy spectra as $\varepsilon_{ns}=-n\omega_{D}+\frac{\omega_{n}}{2}+s\sqrt{\left(\frac{\omega_{D}}{2}+m_{n}\right)^{2}+n\eta^{2}}$ for $n>0$ and $\varepsilon_{0}=-\frac{\omega_{D}}{2}$ with $\omega_{D} =\frac{2d\hbar^{2}}{\ell_{B}^{2}}$. Then, the half-quantization of Hall conductance at $\mu=0$ is destroyed
by a magnetic field as the lowest Landau level becomes field-dependent for $d\ne0$. Nevertheless, the Hall conductance is still half-quantized at $\mu=-\frac{\omega_{D}}{2}$ and integer-quantized when the chemical
potential is inside the gap between two adjacent Landau levels.

\paragraph*{Semi-magnetic topological insulator versus graphene}

Now we come to address the distinctive properties of the Dirac fermions
in a semi-magnetic TI and in graphene. There exists a pair of massless Dirac cones in graphene. It does not break the time-reversal symmetry and there is no Hall conductance for each flavor of massless
Dirac fermions. Thus, no parity anomaly arises. In this case, two linear Dirac cones become a good approximation to describe the massless fermions near the Fermi surface. Each Dirac cone has the half-quantized Hall conductance at a finite field according to Jackiw \citep{Jackiw1984Fractiona}.
If we regard the two flavors of massless Dirac fermions share one lowest Landau level, we can obtain the quantized Hall conductance of $(2\nu+1)e^{2}/h$ as observed experimentally in graphene \citep{Zhang-05nature,Novoselov-05nature}. In a semi-magnetic TI, there exists only a single massless Dirac cone as the time-reversal symmetry is already broken by the magnetically doping on one side. The effective Hamiltonian for the massless surface electrons has to include the symmetry-broken term such that
it avoids violation of the Nielsen-Nonimiya no-go theorem on a lattice. In other words, without the symmetry-broken term, the model could not reflect the correct physics of a single flavor of massless Dirac fermions on a lattice. With the inclusion of the symmetry broken term, we can see that the Hall conductance is one-half in a metallic phase at a zero field and evolves into one or zero in the insulating phase at a strong field. This resolves the puzzle in Beenakker's commentary on experimental measurement in the semimagnetic TIs \citep{Beennakker2022}.

\paragraph*{Summary}

In short, a crossover from one-half quantized Hall conductance in a metallic phase at a zero field to one or zero quantized Hall conductance in an insulating phase at a strong field in the presence of disorders is a clear signature of parity anomaly of a single flavor of massless Dirac fermions on a lattice. For comparison, in graphene and TI films lacking quantum anomaly, the Hall conductance changes from zero to one quantized Hall conductance for a pair of massless Dirac cones in a magnetic field.

\begin{acknowledgments}
This work was supported by the Research Grants Council, University
Grants Committee, Hong Kong under Grant No. C7012-21G and No. 17301823;
the Scientific Research Starting Foundation of University of Electronic
Science and Technology of China under Grant No. Y030232059002011;
and the International Postdoctoral Exchange Fellowship Program under
Grant No. YJ20220059.
\end{acknowledgments}


\begin{thebibliography}{10}
\bibitem{Niemi1983axial}A. J. Niemi and G. W. Semenoff, Axial-Anomaly-Induced
Fermion Fractionization and Effective Gauge-Theory Actions in Odd-Dimensional
Space-Times, Phys. Rev. Lett. \textbf{51}, 2077 (1983).

\bibitem{Redlich1984gauge}A. N. Redlich, Gauge Noninvariance and
Parity Nonconservation of Three-Dimensional Fermions, Phys. Rev. Lett.
\textbf{52}, 18 (1984).

\bibitem{SemenoffPRL1984}G. W. Semenoff, Condensed-Matter Simulation
of a Three-Dimensional Anomaly, Phys. Rev. Lett. \textbf{53}, 2449
(1984).

\bibitem{FradkinPRL1986}E. Fradkin, E. Dagotto, and D. Boyanovsky,
Physical Realization of the Parity Anomaly in Condensed Matter Physics,
Phys. Rev. Lett. \textbf{57}, 2967 (1986).

\bibitem{Matthew-19prb}Matthew F. Lapa, Parity anomaly from the Hamiltonian
point of view, Phys. Rev. B \textbf{99}, 235144 (2019).

\bibitem{Haldane1988Model}F. D. M. Haldane, Model for a Quantum Hall
Effect without Landau Levels: Condensed-Matter Realization of the
\textquotedbl Parity Anomaly\textquotedbl , Phys. Rev. Lett. \textbf{61},
2015 (1988).

\bibitem{Mogi-21np} M. Mogi, Y. Okamura, M. Kawamura, R. Yoshimi,
K. Yasuda, A. Tsukazaki, K. S. Takahashi, T. Morimoto, N. Nagaosa,
M. Kawasaki, Y. Takahashi, and Y. Tokura, Experimental signature of
parity anomaly in semi-magnetic topological insulator, Nat. Phys.
\textbf{18}, 390 (2022).

\bibitem{Jackiw1984Fractiona}R. Jackiw, Fractional charge and zero
modes for planar systems in a magnetic field, Phys. Rev. D \textbf{29},
2375 (1984).

\bibitem{Boyanovsky1986physical}D. Boyanovsky, R. Blankenbecler,
and R. Yahalom, Physical origin of topological mass in 2+ 1 dimensions,
Nucl. Phys. B 270, 483 (1986).

\bibitem{Schakel1991relativistic}A. M. J. Schakel, Relativistic quantum
Hall effect, Phys. Rev. D \textbf{43}, 1428 (1991).

\bibitem{FuB-22xxx}B. Fu, J. Y. Zou, Z. A. Hu, H. W. Wang, and S.
Q. Shen, Quantum anomalous semimetals, npj Quantum Mater. \textbf{7},
94 (2022).

\bibitem{Zou-22prb}J. Y. Zou, B. Fu, H. W. Wang, Z. A. Hu, S. Q.
Shen, Half-quantized Hall effect and power law decay of edge-current
distribution, Phys. Rev. B \textbf{105}, L201106 (2022).

\bibitem{Hu-2022prb}Z. A Hu, H. W. Wang, B. Fu, J. Y. Zou, and S.
Q. Shen, Signature of parity anomaly in the measurement of optical
Hall conductivity in quantum anomalous Hall systems, Phys. Rev. B
\textbf{106}, 035149 (2022).

\bibitem{zou-23prb}J. Y. Zou, R. Chen, B. Fu, H. W. Wang, Z. A. Hu,
and S. Q. Shen, Half-quantized Hall effect at the parity-invariant
Fermi surface, Phys. Rev. B \textbf{107}, 125153 (2023).

\bibitem{Fu2007topological}L. Fu, C. L. Kane, and E. J. Mele, Topological
Insulators in Three Dimensions, Phys. Rev. Lett. \textbf{98}.106803
(2007).

\bibitem{hasan2010colloquium}M. Z. Hasan, and C. L. Kane, Colloquium:
Topological insulators, Rev. Mod. Phys. \textbf{82}, 3045 (2010).

\bibitem{Qi2008topological}X. L. Qi, T. L. Hughes, and S. C. Zhang,
Topological field theory of time-reversal invariant insulators. Phys.
Rev. B 78, 195424 (2008).

\bibitem{shen2012topological}S. Q. Shen, \emph{Topological Insulators},
Springer Series of Solid State Science, Vol. 174 (Springer, Heidelberg,
2012).

\bibitem{XuY-14np} Y. Xu, I. Miotkowski, C. Liu, J. Tian, H. Nam,
N. Alidoust, J. Hu, C. K. Shih, M. Z. Hasan, and Y. P. Chen, Observation
of topological surface state quantum Hall effect in an intrinsic three-dimensional
topological insulator, Nat. Phys. \textbf{10}, 956 (2014).

\bibitem{Chu-11prb}R. L. Chu, J. R. Shi, and S. Q. Shen, Surface
edge state and half-quantized Hall conductance in topological insulators,
Phys. Rev. B \textbf{84}, 085312 (2011).

\bibitem{Koenig2014half}E. J. Koenig, P. M. Ostrovsky, I. V. Protopopov,
I. V. Gornyi, I. S. Burmistrov, and A. D. Mirlin, Half-integer quantum
Hall effect of disordered Dirac fermions at a topological insulator
surface, Phys. Rev. B \textbf{90}, 165435 (2014).

\bibitem{Zhang2017anomalous}S. Zhang, L. Pi, R. Wang, G. Yu, X. -C.
Pan, Z. Wei, J. Zhang, C. Xi, Z. Bai, F. Fei, M. Wang, J. Liao, Y.
Li, X. Wang, F. Song, Y. Zhang, B. Wang, D. Xing and G. Wang, Anomalous
quantization trajectory and parity anomaly in Co cluster decorated
$\mathrm{BiSbTeSe_{2}}$ nanodevices, Nat. Commun. \textbf{8}, 977
(2017).

\bibitem{chang23rmp}C. Z. Chang, C.-X. Liu, and A. H. MacDonald,
Colloquium: Quantum anomalous Hall effect. Rev. Mod. Phys. \textbf{95},
011002(2023).

\bibitem{Novoselov-05nature}K. S. Novoselov, A. K. Geim, S. V. Morozov,
D. Jiang, M. I. Katsnelson, I. V. Grigorieva, S. V. Dubonos, and A.
A. Firsov, Two-dimensional gas of massless Dirac fermions in graphene,
Nature \textbf{438}, 197 (2005).

\bibitem{Zhang-05nature}Y. B. Zhang, Y. W. Tan, H. L. Stormer, and
P. Kim, Experimental observation of the quantum Hall effect and Berry's
phase in graphene, Nature \textbf{438}, 201 (2005).

\bibitem{Neto-09-rmp}A. H. Castro Neto, F. Guinea, N. M. R. Peres,
K. S. Novoselov, and A. K. Geim, The electronic properties of graphene,
Rev. Mod. Phys. \textbf{81}, 109 (2009).

\bibitem{Gusynin-05prl}V. P. Gusynin and S. G. Sharapov, Unconventional
Integer Quantum Hall Effect in Graphene, Phys. Rev. Lett. \textbf{95},
146801(2005).

\bibitem{Yoshimi15nc}R. Yoshimi, A. Tsukazaki, Y. Kozuka, J. Falson,
K.S. Takahashi, J.G. Checkelsky, N. Nagaosa, M. Kawasaki, and Y. Tokura,
Quantum Hall effect on top and bottom surface states of topological
insulator $(\mathrm{Bi}_{1-x}\mathrm{Sb}_{x})_{2}\mathrm{Te}_{3}$
films, Nat. Commun. \textbf{6}, 6627 (2015).

\bibitem{Li17prb}C. Li, B. de Ronde, A. Nikitin, Y. Huang, M. S.
Golden, A. de Visser, and A. Brinkman, Interaction between counter-propagating
quantum hall edge channels in the 3D topological insulator $\mathrm{BiSbTeSe}_{2}$,
Phys. Rev. B \textbf{96}, 195427 (2017).

\bibitem{chong19prl}S. K. Chong, K. B. Han, T. D. Sparks, and V.
V. Deshpande, Tunable Coupling between Surface States of a Three-Dimensional
Topological Insulator in the Quantum Hall Regime, Phys. Rev. Lett.
\textbf{123}, 036804 (2019).

\bibitem{Tokura2019MTI}Y. Tokura, K. Yasuda, and A. Tsukazaki, Magnetic
topological insulators. Nat. Rev. Phys. \textbf{1}, 126 (2019).

\bibitem{liu09prl}Q. Liu, C. X. Liu, C. Xu, X. L. Qi, and S. C. Zhang,
Magnetic Impurities on the Surface of a Topological Insulator, Phys.
Rev. Lett. \textbf{102}, 156603 (2009).

\bibitem{Chen-10science}Y. L. Chen, J. H. Chu, J. G. Analytis, Z.
K. Liu, K. Igarashi, H. H. Kuo, X. L. Qi, S. K. Mo, R. G. Moore, D.
H. Lu, M. Hashimoto, T. Sasagawa, S. C. Zhang, I. R. Fisher, Z. Hussain,
and Z. X. Shen, Massive Dirac fermion on the surface of a magnetically
doped topological insulator, Science \textbf{329}, 659 (2010).

\bibitem{Chang-13science}C. Z. Chang, J. Zhang, X. Feng, J. Shen,
Z. Zhang, M. Guo, K. Li, Y. Ou, P. Wei, L. L. Wang, Z. Q. Ji, Y. Feng,
S. Ji, X. Chen, J. Jia, X. Dai, Z. Fang, S. C. Zhang, K. He, Y. Wang,
L. Lu, X. C. Ma, and Q. K. Xue, Experimental observation of the quantum
anomalous Hall effect in a magnetic topological insulator, Science
\textbf{340}, 167--170 (2013).

\bibitem{Yoshim-15nc}R. Yoshimi, K. Yasuda, A. Tsukazaki, K. S. Takahashi,
N. Nagaosa, M. Kawasaki and Y. Tokura, Quantum Hall states stabilized
in semi-magnetic bilayers of topological insulators, Nat. Commun.
\textbf{6}, 8530 (2015).

\bibitem{lu21prx}R. Lu, H. Sun, S. Kumar, Y. Wang, M. Gu, M. Zeng,
Y.-J. Hao, J. Li, J. Shao, X.-M. Ma et al., Half-Magnetic Topological
Insulator with Magnetization-Induced Dirac Gap at a Selected Surface,
Phys. Rev. X \textbf{11}, 011039 (2021).

\bibitem{Zhou2022prl}H. Zhou, H. Li, D. H. Xu, C. Z. Chen, Q. F.
Sun, X. C. Xie, Transport Theory of Half-Quantized Hall Conductance
in a Semimagnetic Topological Insulator, Phys. Rev. Lett.\textbf{
129}, 096601 (2022).

\bibitem{Gong2023nsr}M. Gong, H. Liu, H. Jiang, C. Z. Chen, X. C.
Xie, Half-Quantized Helical Hinge Currents in Axion Insulators, Natl.
Sci. Rev. nwad025 (2023).

\bibitem{Beennakker2022}C. Beenakker, Anomalous quantum anomalous
Hall effect, J. Club Cond. Matter October 02 (2022). \href{https://www.condmatjclub.org/?p=4688}{https://www.condmatjclub.org/?p=4688}

\bibitem{Shan2010njp}W. Y. Shan, H. Z. Lu and S. Q. Shen, Effective
continuous model for surface states and thin films of three-dimensional
topological insulators, New J. Phys. \textbf{12}, 043048 (2010).

\bibitem{Nielsen1981PLB}H. B. Nielsen and M. Ninomiya, A no-go theorem
for regulatizing chiral fermions, Phys. Lett. B \textbf{105}, 219
(1981).

\bibitem{Wilson-75}K. G. Wilson, \emph{New Phenomena in Subnuclear
Physics}, ed. A. Zichichi (New York, Plenum, 1975).

\bibitem{Rothe}H. J. Rothe, \emph{Lattice gauge theories: an introduction},
3rd ed. (World Scientific, Singapore, 2005).

\bibitem{shen-05prb}S. Q. Shen, Y. J. Bao, M. Ma, X. C. Xie, and
F. C. Zhang, Resonant spin Hall conductance in quantum Hall systems
lacking bulk and structural inversion symmetry, Phys. Rev. B \textbf{71},
155316 (2005).

\bibitem{streda-1982}P. Streda, Quantised Hall effect in a two-dimensional
periodic potential, J. Phys. C: Solid State Phys. \textbf{15}, L1299
(1982).

\bibitem{wang-19prb}H. W. Wang, B. Fu, and S. Q. Shen, Intrinsic
magnetoresistance in three-dimensional Dirac materials with low carrier
density, Phys. Rev. B \textbf{98}, 081202(R) (2018).

\bibitem{note-SI}See Supplemental Material at {[}URL to be added
by publisher{]} for details of (Sec. S1) Hall conductance in the zero
magnetic field, (Sec. S2) Hall conductance in the finite magnetic
field, (Sec. S3) longitudinal conductance, and (Sec. S4) calculation
of spectral asymmetry $\eta_{H}$, which includes Refs. \citep{shen-05prb,streda-1982,wang-19prb,niemi-eta,B=0000F6ttcher2019survival}.

\bibitem{APS-index}M. F. Atiyah, V. K. Patodi, and I. M. Singer,
in \emph{Mathematical Proceedings of the Cambridge Philosophical Society}
(Cambridge University, Cambridge, England, 1975), Vol. \textbf{77},
pp. 43--69.

\bibitem{niemi-eta}A. J. Niemi, Topological solitons
in a hot and dense fermi gas, Nucl. Phys. B \textbf{251}, 155 (1985).

\bibitem{B=0000F6ttcher2019survival}J. B{\"o}ttcher, C. Tutschku, L.
W. Molenkamp, and E. e M. Hankiewicz, Survival of the Quantum
Anomalous Hall Effect in Orbital Magnetic Fields as a Consequence
of the Parity Anomaly, Phys. Rev. Lett. \textbf{123}, 226602 (2019).

\bibitem{B=0000F6ttcher2020fate}J. B{\"o}ttcher, C. Tutschku, and E.
M Hankiewicz, Fate of quantum anomalous Hall effect in the presence
of external magnetic fields and particle-hole asymmetry, Phys. Rev.
B \textbf{101}, 195433 (2020).
\end{thebibliography}
\end{document}